# Deterministic control of antiferromagnetic domain walls by circular phonons

Vladislav Bilyk[1], Nikolai Khokhlov[1], Viktoriia Radovskaia[1], Peter Kim[1], Pietro Diona[2,7], Ravi Kaushik[2,8], Luca Maranzana[2,8], Takeshi Hayashida[3,1], Sergey Artyukhin[2], Edwin Hang Tong Teo[4,5], Apoorva Chaturvedi[6], Carl S. Davies[3,1], Andrei Kirilyuk[3,1], Alexey Kimel[1]; Dmytro Afanasiev[1]

[1] Institute for Molecules and Materials, Radboud University, Nijmegen, 6525 AJ, Netherlands

[2] Quantum Materials Theory, Italian Institute of Technology, Via Morego 30, 16163 Genova, Italy

[3] HFML-FELIX, Radboud University, Toernooiveld 7, 6525 ED Nijmegen, The Netherlands

[4]School of Electrical and Electronic Engineering, Nanyang Technological University, Singapore 639798, Singapore

[5]School of Materials Science and Engineering, Nanyang Technological University, Singapore 639798, Singapore

[6]Temasek Laboratories, Nanyang Technological University, 50 Nanyang Drive, 637553, Singapore

[7]Nanoscience, Scuola Normale Superiore, Piazza dei Cavalieri 7, Pisa, Italy

[8]Department of Physics, University of Genoa, Via Dodecaneso 33, Genoa, Italy

In an antiferromagnet, areas with different orientations of the antiferromagnetic Néel vector, which plays the role of the order parameter, can coexist, thus forming antiferromagnetic domains separated by domain walls. This suggests that antiferromagnets can possess the functionalities of magnetic storage media. Whether one can benefit from these functionalities crucially depends on the availability of efficient means for deterministic control of antiferromagnetic domain walls. Here, we demonstrate an approach to deterministically control domain walls in the van der Waals antiferromagnet Ni-doped $MnPS_3$ via dynamic engineering of the crystal lattice. By resonantly exciting a pair of nearly degenerate orthogonal infrared-active $A_u$ and $B_u$ phonon modes with circularly polarized mid-infrared light, we induce helicity-dependent reconfiguration of 180° antiferromagnetic domains, providing evidence for a phonon-induced effective field conjugate to the Néel order parameter. The domain kinetics exhibit a pronounced helicity asymmetry governed by the interplay between domain-wall elasticity and defect-mediated pinning, enabling small phonon-driven perturbations to accumulate into stable domain transformations. Our results establish dynamic lattice control and defect engineering as means for deterministic control of antiferromagnetic order.

# Introduction

The ability to reversibly and deterministically control magnetic order is a centerpiece of modern magnetic information technologies[1,2], with growing interest shifting from ferromagnetic to antiferromagnetic media due to their potential for ultrafast operation and high information density[3,4]. Central to this effort is the manipulation of magnetic domains - regions of uniform magnetic order separated by domain walls. The controlled motion of these walls enables transitions between distinct magnetic states and underpins information storage and processing[5-7]. While ferromagnetic domains can be deterministically manipulated by controlling the sign of an external magnetic field, antiferromagnetic domains may carry no macroscopic magnetic moment, making such deterministic control considerably more challenging. Moreover, unlike in ferromagnets, where long-range dipolar interactions among spins play a dominant role in defining the domain structure, the origin and stability of antiferromagnetic domains cannot be understood taking spin system alone. It is believed that the spatial distribution of antiferromagnetic domains is largely governed by magnetoelastic interactions, while structural defects and inhomogeneous strain generate a pinning landscape that defines the motion of the antiferromagnetic domain walls[8-12].

Strain engineering has recently emerged as a promising strategy for controlling antiferromagnetic domains and domain walls. By modifying lattice parameters and local crystal symmetries, one can tune magnetic anisotropy, exchange interactions, and domain-wall pinning barriers that govern the stability of antiferromagnetic domain textures[5,13-16]. Both static mechanical strain[17-20] and optically-induced dynamic strain[21-23] have been shown to drive spin-reorientation transitions and reconfigure antiferromagnetic domain textures. However, on-demand deterministic and thus reversible switching of the antiferromagnetic order parameter, the Néel vector **L**, has yet to be realized. While strain can modify the magnetic free energy and reconfigure domains, it is even under time reversal and therefore does not couple directly to the time-reversal-odd Néel vector as its thermodynamic conjugate field. Achieving such control instead requires a stimulus that is thermodynamically conjugate to **L** and thus shares its symmetry properties with respect to time-reversal and space inversion[3].

A promising avenue to realize such a stimulus is offered by circular phonons[24,25], in which the crystal lattice is microscopically driven into chiral motion by resonant excitation with circularly polarized terahertz or mid-infrared light. Under these conditions, the coherent ionic displacements acquire a well-defined angular momentum determined by the helicity of the driving field, dynamically breaking time-reversal symmetry[26-29]. This chiral lattice motion can couple to spins and thereby deliver a phononic analogue of the inverse Faraday effect[16,30-32] with theoretically predicted effective magnetic fields reaching 100 T [33]. Recent experiments have suggested that circular lattice excitations are able to induce ultrafast helicity-dependent transient magnetization in paramagnets[34] and to enable coherent control of spin dynamics, including phase control of spin precession in canted antiferromagnets[24] and deterministic switching in ferrimagnets[23]. Here, we explore whether resonantly excited circular phonons can provide deterministic and reversible control over antiferromagnetic domain walls. We show that resonant excitation of chiral lattice motion with circularly polarized mid-infrared light realizes precisely such a symmetry-matched stimulus, enabling direct control of the Néel order parameter and its domain texture (Fig. 1a).

**Sample and experimental setup**

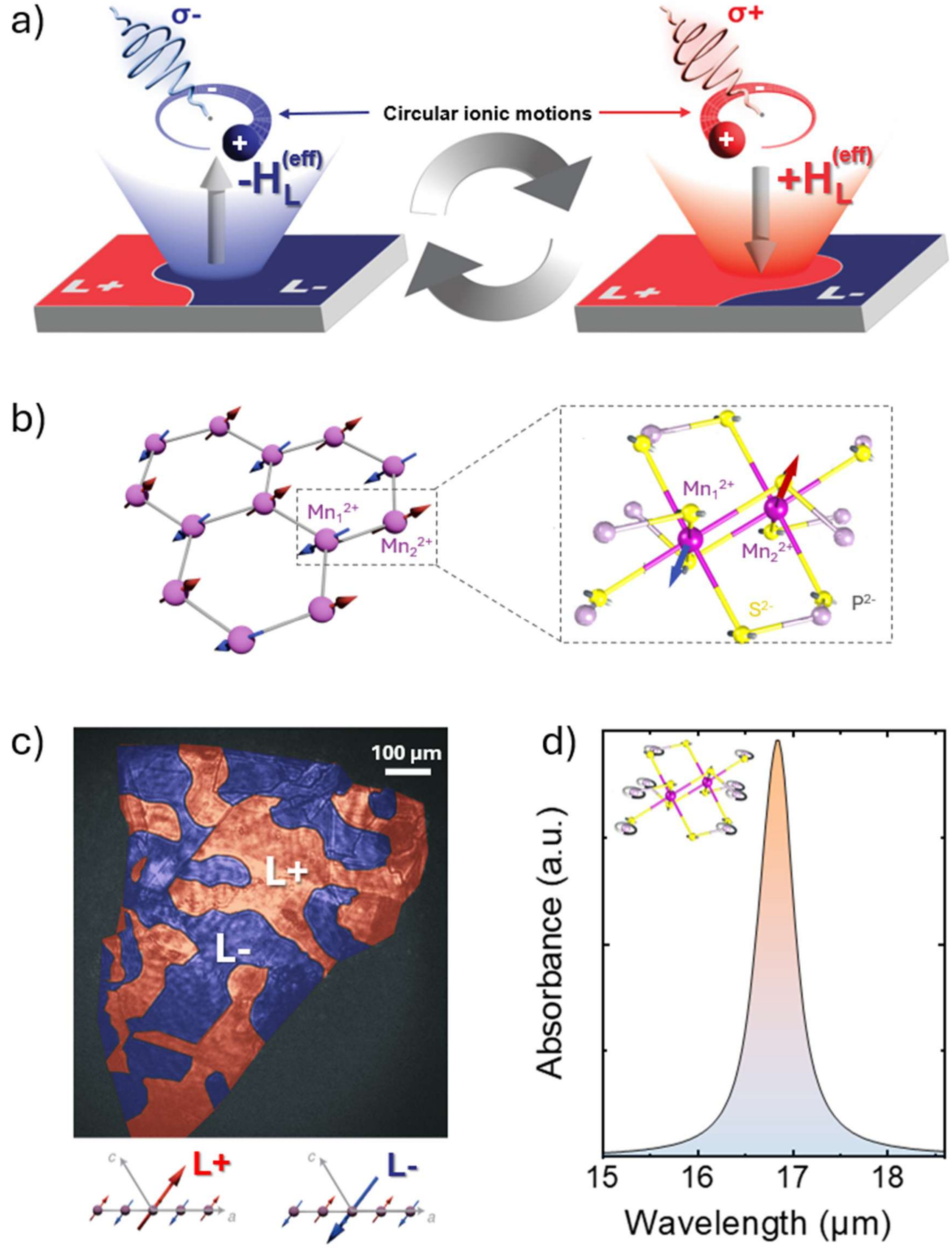

**Figure 1. Experimental concept and material platform. (a)** Laser-induced coherent circularly polarized lattice vibrations act as a helicity-dependent driving field for reversible reconfiguration of the antiferromagnetic (AFM) domain structure, determining the final state. **(b) Left**: magnetic structure of a single layer of $MnPS_3$:Ni, illustrating its *PT*-symmetric AFM order, **right**: crystallographic structure. **(c) Top:** SHG microscopy image of a $MnPS_3$:Ni flake revealing the AFM domain structure, **bottom**: schematics of oppositely oriented Néel domains**. (d)** Mid-infrared absorption spectrum of the $MnPS_3$:Ni sample, highlighting absorption associated with the highest-energy infrared-active nearly degenerate $A_u+B_u$ phonon modes leading to a circular motion of $(P_2S_6)^{4-}$ ligand complex (see inset).

To explore lattice-driven deterministic control of antiferromagnetic domain walls, we chose $MnPS_3$, a layered van der Waals antiferromagnet with a low-symmetry monoclinic crystal structure (space group $C2/m$). Below its Néel temperature, $MnPS_3$ develops collinear antiferromagnetic order on a honeycomb lattice that can be effectively viewed as two equivalent ferromagnetic sublattices with opposite magnetizations, $\mathbf{M}_1$ and $\mathbf{M}_2$. As their exact compensation ($\mathbf{M}_1 + \mathbf{M}_2 = 0$) results in zero net magnetization, the magnetic order can instead be described by the Néel vector $\mathbf{L} = \mathbf{M}_2 - \mathbf{M}_1$. In $MnPS_3$ the **L-**vector is oriented predominantly out of the basal plane, with a slight canting of a few degrees toward the plane induced by weak easy-plane anisotropy[35]. Owing to the invariance of the magnetic free energy under time reversal, the two states $\mathbf{L}\pm$ are energetically degenerate. As a consequence, time-reversed 180° antiferromagnetic domains form which are separated by domain walls. Although the crystal lattice itself is centrosymmetric, such antiferromagnetic spin arrangement breaks both spatial inversion ($P$) and time-reversal ($T$) symmetries individually, while preserving their combined $PT$ symmetry (Fig. 1b).

Owing to broken inversion symmetry, the antiferromagnetic domains can be directly visualized via a second-order nonlinear optical response, which we refer to as second harmonic generation (SHG)[36-38]. Although SHG does not directly distinguish the sign of the Néel vector and therefore yields the same intensity from the **L+** and **L-** domains, it enables a clear visualization of their boundaries. At the domain walls, where the Néel vector continuously rotates between adjacent domains, the SHG signal is strongly suppressed, enabling the domain texture to be reconstructed from the resulting network of dark boundaries (see Supplementary Figure S1). To enhance the SHG signal and thus to improve the visibility of the domain texture, we employ lightly Ni-doped $MnPS_3$ ($MnPS_3$:Ni)[39]. The Ni substitution preserves the Néel-type antiferromagnetic order below $T_N$=70 K while canting the Néel vector slightly away from the surface normal, thereby generating a finite in-plane component that substantially enhances the SHG contrast[20]. Figure 1c shows an SHG image of a thin (~10 µm) $MnPS_3$:Ni flake exfoliated onto a transparent $CaF_2$ substrate and cooled down to $T$=6 K, revealing a labyrinth-like pattern of time-reversed Néel domains, labeled "red" and "blue" and corresponding to the states with mutually opposite orientations of the antiferromagnetic Néel vector **L+** and **L-**, respectively. Our measurements further show that the domain configuration is exceptionally robust to temperature. The Néel pattern remains unchanged up to $T_N$, and can be only changed upon thermal cycling across $T_N$ (see Supplementary Figure S2).

To realize deterministic control of the Néel domain configuration in $MnPS_3$:Ni, we exploit the unique features of its infrared-active phonon spectrum. Our DFT calculations reveal several nearly degenerate pairs of orthogonally polarized $A_u$ and $B_u$ infrared-active phonons. Resonant excitation of such a phonon pair with circularly polarized light coherently drives both eigenmodes, generating circular lattice motion with well-defined handedness. We focus on the highest-energy $A_u$+$B_u$ pair at 16.5 µm (≈75 meV), see Fig. 1d, which exhibits the largest oscillator strengths across entire phonon spectrum (see Supplementary Figure S3). The corresponding lattice motion is dominated by the internal vibrations of the $(P_2S_6)^{4-}$ ligand complex, with only a minor contribution from the $Mn^{2+}$ ions [40,41]. In particular, the coherent superposition of the two phonons produces a pronounced circular displacement of the phosphorus atoms (Fig. 1b), which directly modulates the crystal field surrounding the $Mn^{2+}$ sites. Through spin–orbit coupling, this chiral crystal-field modulation is expected to generate a helicity-dependent effective field acting on the Néel order, analogous to the phononic inverse Faraday effect[16,30-32].

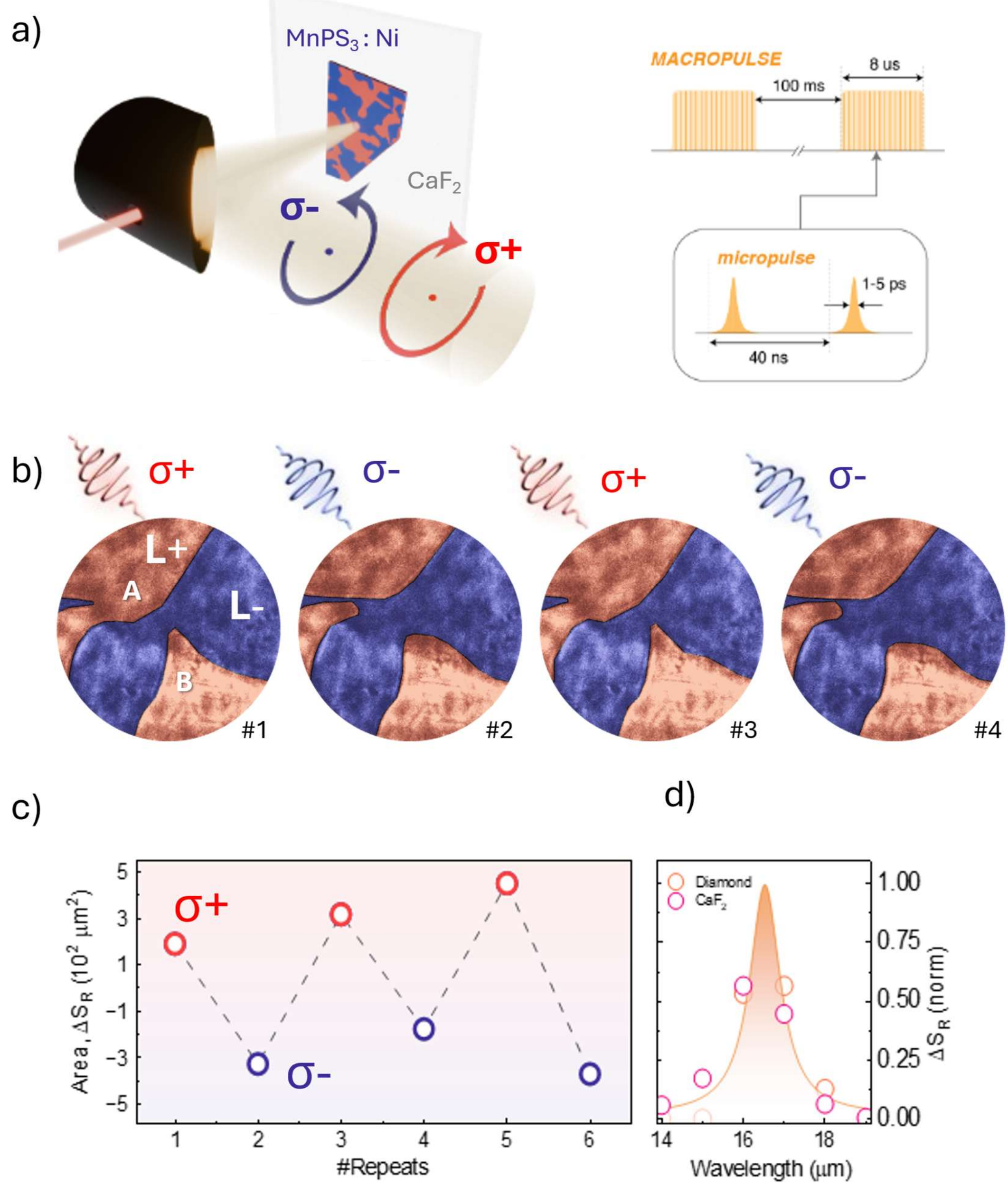

**Figure 2. Helicity-dependent phononic reconfiguration of Néel domains**. **(a)** Schematic illustration of the sample under mid-infrared excitation with opposite circular helicities, together with the temporal structure of the MIR macropulses delivered by FELIX (right). **(b)** Sequence of SHG images acquired after exposure to right- (σ+) and left-circularly (σ-) polarized MIR pulses, showing reversible toggling between a pair of domain reconfiguration. Each image corresponds to a field of view of 200 μm in diameter. **(c)** Fraction of the red-domain area as a function of exposure to left- and right-circularly polarized MIR pulses, demonstrating helicity-dependent and reversible control of the domain population. **(d)** Normalized helicity-induced change in the red-domain area as a function of excitation wavelength for $MnPS_3$:Ni exfoliated onto diamond and $CaF_2$ substrates.

To resonantly excite the targeted phonon modes, we employed the Free Electron Lasers for Infrared eXperiments (FELIX) in Nijmegen, the Netherlands[42]. FELIX delivers narrow-band, wavelength-tunable mid-infrared (MIR) radiation as a train of linearly polarized macropulses at a repetition rate of 10 Hz. Each macropulse consists of approximately 200 micropulses with a duration of ~1 ps (Fig. 2a). The polarization of the MIR beam was converted to left- or right-handed circular polarization using a wavelength-tunable zero-order quarter-wave plate, while maintaining the beam position and spatial alignment at the sample, see Methods. By focusing circularly polarized FELIX pulse sequences onto the $MnPS_3$:Ni surface and imaging the evolution of the antiferromagnetic domain texture, we directly reveal how resonant driving circular lattice motion couples to and reconfigures the Néel order.

**Helicity-dependent phononic switching of the Néel vector**

We first investigated how exposure to multiple macropulses of circularly polarized FELIX MIR radiation affects the antiferromagnetic domains. The total irradiation time was set to about 10 s, corresponding to cumulative pumping by more than 100 macropulses ($10^4$ picosecond-long micropulses). Our experiments show that the cumulative exposure to such prolonged circular lattice excitation changes the underlying antiferromagnetic domain pattern. The resulting configuration remains stable after the excitation is turned off, as well as under continued irradiation with the same helicity, indicating the settling of a new stable spatially inhomogeneous state. The leftmost panel of Figure 2b zooms in on one such stable configuration obtained under right-circularly polarized excitation ($\sigma^+$), herein referred to as #1. Remarkably, irradiating this state with left-circularly polarized ($\sigma^-$) MIR radiation transforms the domain pattern into a new configuration, denoted #2. The resulting state is again stable, showing no further evolution under continuous $\sigma^-$ irradiation and remaining unchanged after the excitation is turned off. Importantly, not all antiferromagnetic domain walls within the illuminated region respond equally to the MIR radiation. Within the studied area, the changes are most pronounced in the regions marked “A” and “B”. In region “A”, the domain wall of one of the initially separated “blue” $\mathbf{L}^-$ domains advances across the adjacent “red” **L+** domain, causing the **L-** domain to elongate and eventually form a bridge to the neighbouring “blue” $\mathbf{L}^-$ domain. In region “B”, the domain wall recedes, causing a sharp protrusion of the red **L+** domain into the surrounding **L**- domain to retract and become progressively smoother. Despite the complexity, the net effect of the $\sigma^-$ excitation can be seen as a systematic expansion of the “blue” **L**- domains at the expense of the “red” domains within the considered region. Strikingly, reversing the helicity to $\sigma^+$ (Shot #3) largely restores the preceding domain configuration, effectively undoing the modifications induced by the prior exposure and reducing the fraction of the “blue” **L**- domains. Further experiments reveal that the same helicity-sensitive regions reproducibly switch between these two configurations upon MIR helicity reversal, demonstrating high-fidelity deterministic and reversible control of a domain texture (Shots #1-#4). This robustness is further quantified in Fig. 2c, where we introduce $\Delta S_{\mathrm{R}}$ parameter, which represents the change in the “red”-domain area relative to the unperturbed domain configuration prior to MIR excitation. The reproducible increase and decrease in $\Delta S_{\mathrm{R}}$ upon successive helicity reversals provides a quantitative measure of the reversible Néel-domain area modulation. Such behaviour is consistent with the emergence of a phonon-helicity-dependent effective field, $\mathbf{H}_{\mathrm{L}}^{(\mathrm{eff})}$, which is thermodynamically conjugate to the Néel vector and thus directly acting on it such that $\mathbf{H}_{\mathrm{L}}^{(\mathrm{eff})} = -\frac{\partial \Phi_{\mathrm{AFM}}}{\partial \mathbf{L}}$, where $\Phi_{\mathrm{AFM}}$ is the thermodynamic potential of the antiferromagnet. Similarly to the Néel vector in the antiferromagnet the effective field $\mathbf{H}_{\mathrm{L}}^{(\mathrm{eff})}$ must change sign under both time-reversal and spatial inversion, thus exhibiting the signatures of a chiral excitation. This field breaks the degeneracy between states with opposite orientations of the antiferromagnetic Néel vector and can therefore move domain wall deterministically and reversibly, enabling expansion of the domains with the energetically favoured Néel vector orientation.

We also note that the degree of helicity-dependent control is not uniform across the flake: certain regions are markedly more susceptible to the phononic pumping than others. This is consistent with the observation that the demonstrated reconfiguration is confined to specific, highly susceptible regions of the domain texture and does not simply follow the Gaussian profile of the excitation beam. Such selectivity is suggestive that the domain-wall motion underlying the expansion of Néel domains is strongly governed by pinning centers. The pinning centers can originate from crystal defects and locally varying strain fields, etc., all of which are commonly encountered in exfoliated van der Waals magnets[43,44] and create spatial variations in the energy barrier for domain-wall motion. Consequently, the effective field acting on the Néel order must locally assist the domain wall in overcoming pinning barriers, initiating directional domain-wall motion between pinning centers. The subsequent propagation is limited when the domain wall encounters stronger pinning sites, resulting in spatially selective switching within the illuminated region.

To confirm the essential role of the infrared-active phonon modes of $MnPS_3$ in the helicity-dependent, deterministic, and reversible control of the antiferromagnetic domain pattern, we performed wavelength-dependent MIR excitation measurements. By tuning the pump wavelength, we again quantified the area $\Delta S_R$ of the reversibly switched domains. The switched domain area exhibits a pronounced maximum at 16.5 µm, coinciding with the $A_u$+$B_u$ phonon resonance, and decreases rapidly away from the resonance, closely following its absorption profile (Fig. 2d). The same wavelength dependence was reproduced on two different substrates, $CaF_2$ and diamond (see also Supplementary Figure S4), confirming that the effect is intrinsic to the studied $MnPS_3$:Ni system. Together with the helicity dependence, the resonant response confirms the infrared-active phonons as the origin of the effective field that lifts the degeneracy between antiferromagnetic states with opposite Néel vector orientations.

**Temperature limits of the reversible phononic control**

To evaluate the robustness of the helicity-dependent reversible switching, we examined the temperature dependence of the phononic control up to the Néel temperature $T_N$. Figure 3 shows the effect of circularly polarized phonon pumping on a large, isolated antiferromagnetic Néel domain wall pinned at an elongated structural defect, where the domain boundary coincides with a step in the number of van der Waals layers, see left panel in Fig. 3a. At the lowest accessible temperature ($T$=6 K), circularly polarized phonon pumping induces a pronounced reconfiguration of the domain structure. Consistent with our previous observations, domain reconfiguration is observed not only within the irradiated region but also at locations well beyond the pump spot, see central panel in Fig. 3a. This behavior demonstrates that the stable configuration of the domain wall is defined not only by the pinning centers. While randomly distributed pinning centers aim to bend the wall, the elastic energy of it resists the bending thus causing domain wall to also move outside the irradiated area.

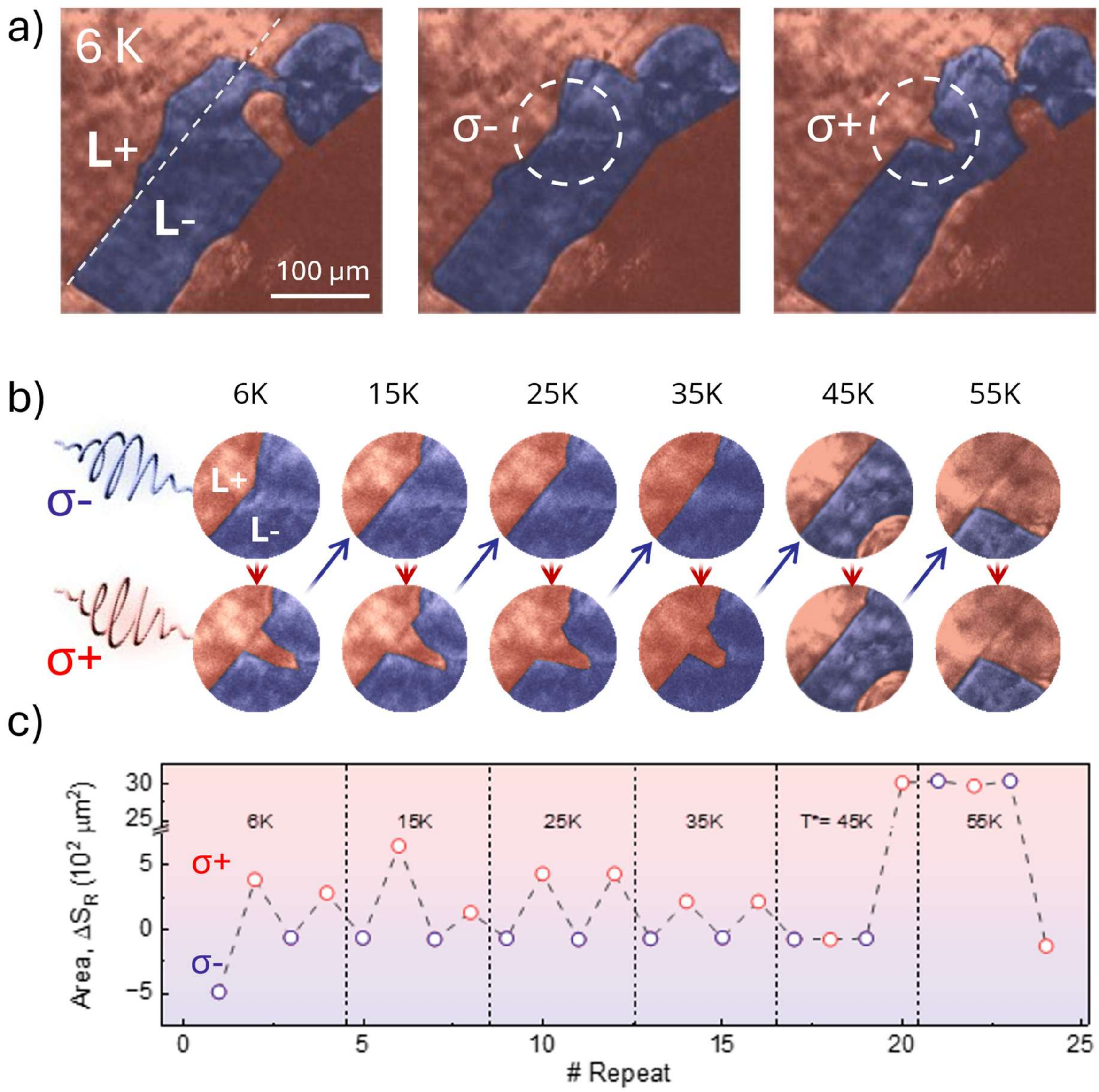

**Figure 3. Temperature limits of the reversible phononic switching (a)** Microscopical SHG image of the domain pattern taken before any exposure and immediately after exposure with left-circular pulse. The elliptical area represents position and spatial profile of mid-IR pulses. Dashed line in left panel denotes crystal defect. **(b)** Zoomed-in SHG images of the same region after excitation with left- and right-circularly polarized MIR pulses at different temperatures, highlighting the temperature dependence of the helicity-induced domain reconfiguration. Each image corresponds to a field of view of 100 μm in diameter. The arrows show a sequence of the measurements. **(c)** "Red"-domain area fraction as a function of temperature following excitation with left- and right-circularly polarized MIR pulses, demonstrating a strong temperature dependence of the helicity-controlled domain switching.

To quantify the reversible domain reconfiguration, we focus on a helicity-sensitive region at the center of the pump spot, where the domain wall locally depins and forms a well-defined indentation into the neighbouring **L**- "blue" domain (Fig. 3b). In this region, we verify that reversing the pump helicity restores the original domain configuration and demonstrate that this response is reproducible over multiple switching cycles (see Fig. 3b and Supplementary Figure S5). We further track the temperature dependence of this behaviour (Fig. 3c) by quantifying the indentation area $\Delta S_R$ under repeated helicity reversals. After an initial exposure to $\sigma^-$ at $T$=6 K establishes a stable domain configuration, the system subsequently exhibits robust helicity-driven reversibility of the domain pattern in the range from 6 K up to $T^*\approx$45 K. Above $T^*$, however, this reversibility breaks down, and further changes in helicity have only a weak effect on the Néel domains.

**Kinetics of single- and multi-pulse Néel-domain reconfigurations**

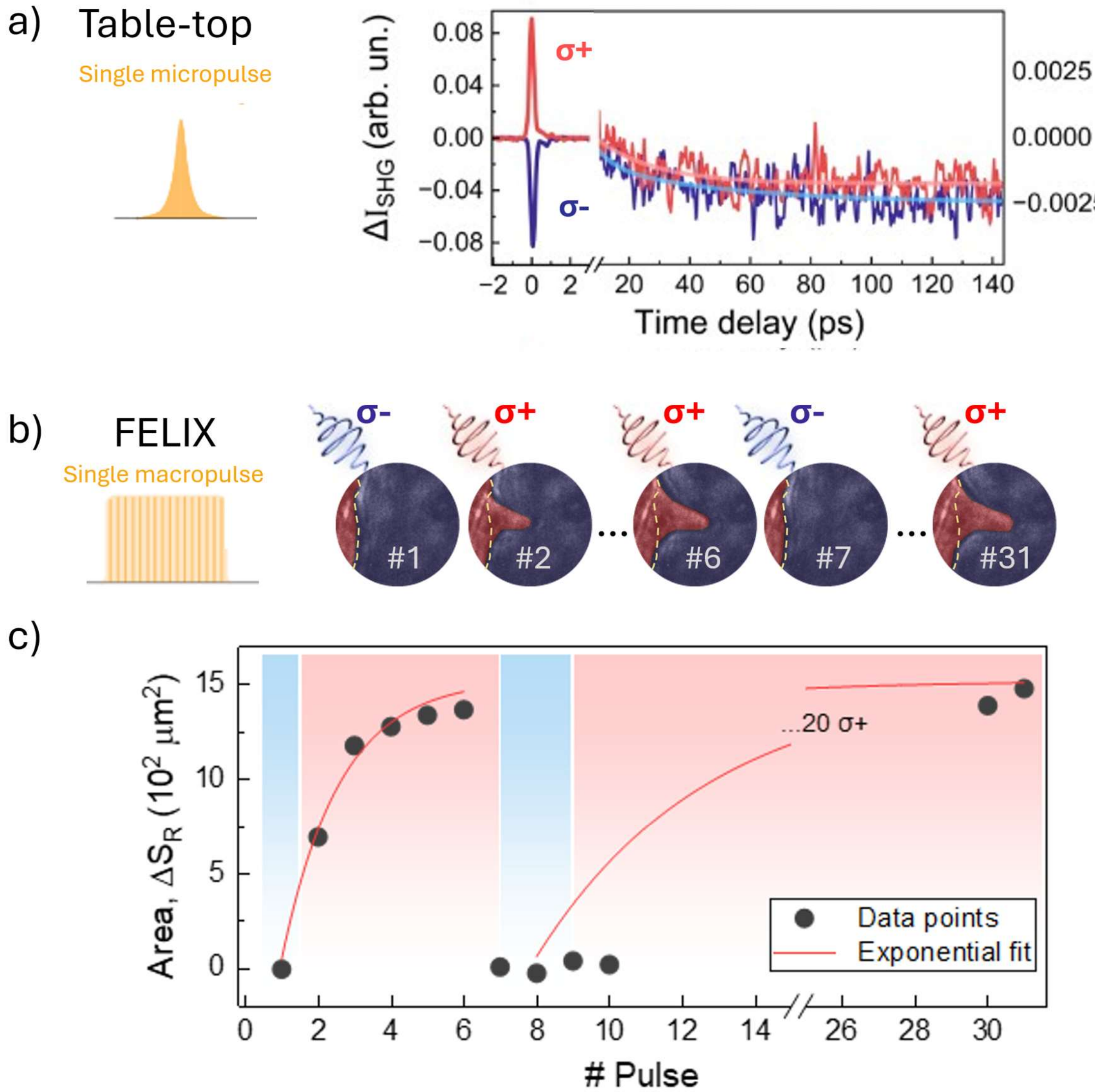

**Figure 4. Multiscale dynamics of helicity-dependent phononic switching. (a)** Left: Schematic illustration of micropulse excitation. Right: Time-resolved SHG dynamics following micropulse excitation with opposite circular helicities, showing helicity-dependent response. **(b)** Left: Schematic illustration of macropulse excitation. Right: SHG microscopy images showing gradual growth and sudden collapse of the red-domain fraction induced by successive action of a single MIR macropulse. The yellow dashed line indicates the initial domain-wall position after excitation with left-circular pulse and is used as a reference to highlight the evolution of the domain configuration. Each image corresponds to a field of view of 100 µm in diameter. **(c)** Red-domain area fraction as a function of successive MIR macropulses, demonstrating cumulative helicity-dependent domain growth.

To reveal the underlying kinetics of the antiferromagnetic order to a single circularly-polarized picosecond MIR pulse, we perform time-resolved stroboscopic SHG measurements with femtosecond temporal resolution (see Methods). Without spatial resolution, these measurements probe the ultrafast response of the antiferromagnet as a whole. The time-resolved-SHG signal shows a pronounced helicity-dependent peak during the temporal overlap of the pump and probe pulses, followed by a helicity-independent change in the SHG intensity with a characteristic timescale of about 50 ps (Fig. 4a). As the magnetic SHG is sensitive only to the magnitude, but not the sign, of the Néel vector, the initial helicity-dependent response is unlikely to originate from a reorientation of the antiferromagnetic order. The subsequent change in SHG intensity, in contrast, is rather consistent with a transient change in the magnitude of the antiferromagnetic order parameter, $|\mathbf{L}|$. Indeed, the characteristic timescale is comparable to the previously reported laser-induced dynamics in $MnPS_3$[36] and can likely be attributed to spin-lattice relaxation. Importantly, the magnitude of this pump-induced change is a small percentage of the static SHG signal, indicating that a single pump pulse produces only a small perturbation of the antiferromagnetic state. The helicity-dependent domain reconfiguration observed in the spatially-resolved measurements therefore requires the cumulative action of multiple pump pulses, analogous to the recently reported phononic switching via ultrafast Barnett effect in GdFeCo [45].

To further examine how the initial response to single-pulse excitation builds-up, we track the evolution of the antiferromagnetic domains as function of the number of FELIX macropulses. By applying one macropulse at a time and imaging the domain configuration after each pulse, we directly track the cumulative effect of the MIR light on the antiferromagnetic domains (see Fig. 4b). Prepared by prolonged exposure to multiple $\sigma^-$ macropulses (Shot #1), the initial domain configuration consists of a pair of coexisting **L+** and **L-** domains separated by domain walls, with the **L-** ("blue") domain occupying the larger area. Applying a single $\sigma^+$ macropulse (Shot #2) to this initial state promotes a substantial reconfiguration, giving rise to an "icicle-like" protrusion extending from the domain boundary, similar to that seen in the temperature-dependent measurements (Fig. 3b). Subsequent $\sigma^-$ (Shots #3-#6) macropulses do not qualitatively modify the shape of the feature, but progressively increase its length, thereby enlarging the area of the "red" domain. The growth saturates after approximately four pulses, indicating that the phonon-driven domain reconfiguration is cumulative, spatially localized, and saturable. This evolution is quantified in Fig. 4c, which shows the change in domain area $\Delta S_{\mathrm{R}}$ as a function of pulse number. Remarkably, while the saturated formation of the icicle-like feature requires multiple macropulses of $\sigma^+$ helicity, a single macropulse of $\sigma^-$ helicity is sufficient to reverse the domain modification, restoring the configuration close to the initial $\sigma^-$ -prepared state (Shot #7-#8). Reapplying a sequence of $\sigma^+$ (Shot #9-#31) pulses then progressively rebuilds the protrusion.

To elucidate a possible mechanism underlying this asymmetry, we performed simulations of a single antiferromagnetic domain wall evolving in an inhomogeneous pinning landscape (for details see Supplementary Information S6 and S7). The model considers a domain wall pinned by strong defects, with a locally defect-free region introduced at the center of the simulated geometry. In addition, weaker pinning centers are uniformly distributed throughout the system, representing the background disorder landscape that stabilizes local domain-wall displacements. The effect of circularly polarized phonons was modeled by introducing a sequence of unipolar pulses representing a phonon-helicity-dependent effective field $\mathbf{H}_{\mathrm{L}}^{(\mathrm{eff})}$, acting on the Néel vector along the sample normal. When $\mathbf{H}_{\mathrm{L}}^{(\mathrm{eff})} > 0$, the domain wall is driven outward, deforming its central section within the defect-free region, while the surrounding segments remain largely immobilized by pinning (Fig. 5a). The central defect-free region enables domain-wall motion, whereas the surrounding distributed pinning centers progressively capture the displaced wall, stabilizing each incremental displacement and preventing complete relaxation between pulses. Consequently, successive pulses accumulate the deformation, leading to the characteristic icicle-shaped domain. The growth saturates after approximately seven pulses, when the driving force exerted by the effective field is balanced by the restoring forces associated with domain-wall elasticity and pinning. Once the protrusion has fully developed, a single pulse of opposite polarity $\mathbf{H}_\mathrm{L}^{(\mathrm{eff})} < 0$ produces the opposite behaviour. In this case, the field-induced force acts in the same direction as the restoring force arising from the domain-wall tension, causing the protrusion to collapse rapidly and the domain wall to return close to its initial configuration after a single pulse (rightmost snapshot in Fig. 5a). These simulated dynamics reproduce the experimentally observed asymmetric response remarkably well (compare Fig. 5b and Fig. 4e).

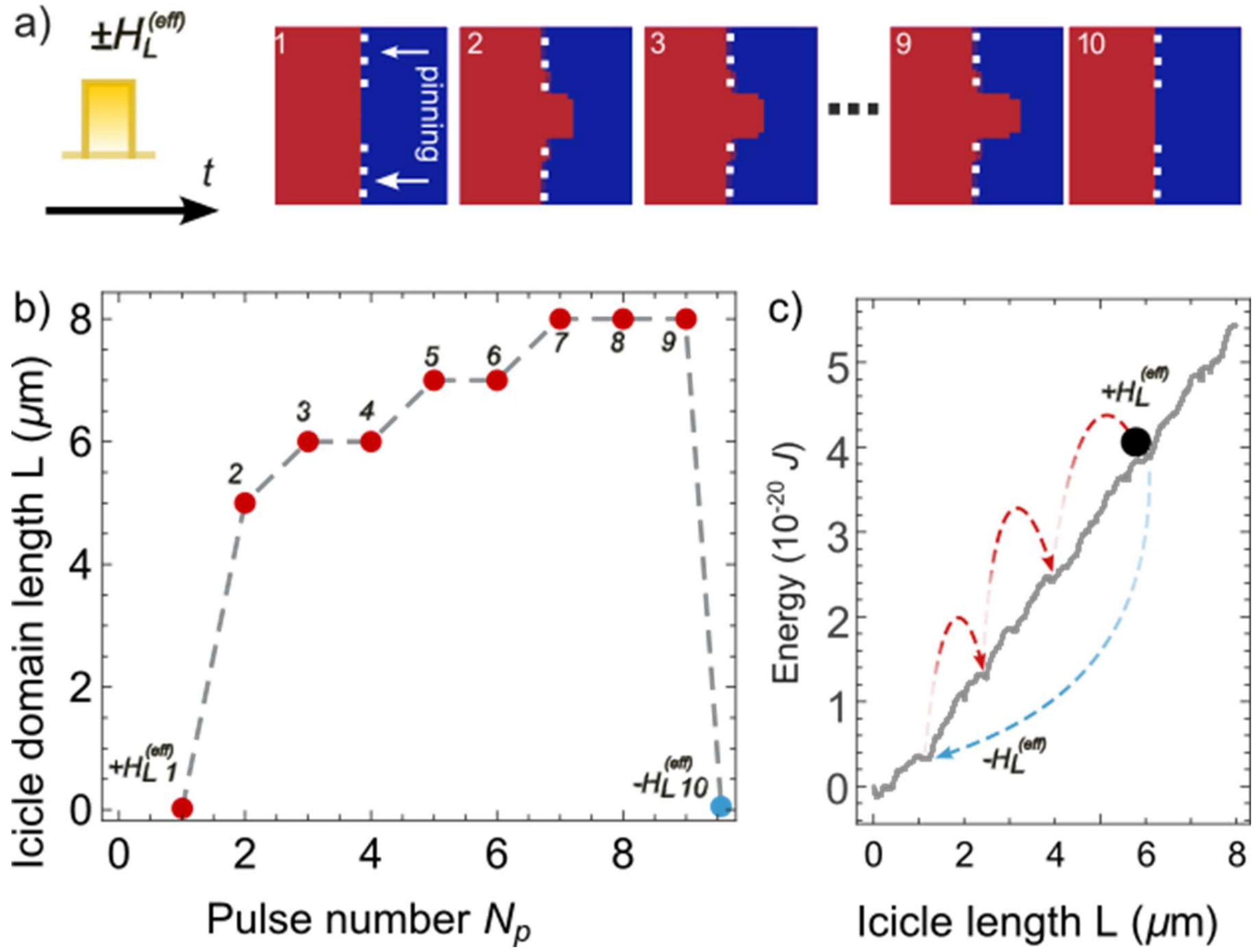

**Figure 5. Micromagnetic simulations of phonon-driven domain reconfiguration in the presence of pinning.** (**a**) Evolution of the antiferromagnetic domain texture under successive applications of the effective phononic field $\mathbf{H}_\mathrm{L}^{(\mathrm{eff})}$ acting on the Néel order parameter in the presence of defect-mediated pinning. (**b**) Simulated change in the switched domain area as a function of the number of excitation pulses. (**c**) Calculated pinning potential experienced by the propagating domain wall, showing the evolution of the characteristic icicle-like domain protrusion with increasing pulse number. Arrows indicate the direction of domain-wall motion under successive fields of opposite sign ($+\mathbf{H}_\mathrm{L}^{(\mathrm{eff})}$ and $-\mathbf{H}_\mathrm{L}^{(\mathrm{eff})}$).

This behaviour can be understood in terms of the effective energy landscape governing the motion of the pinned domain wall (Fig. 5c). Pinning centers create an energy ladder of local minima separated by barriers, corresponding to successive metastable configurations of the propagating wall. Each excitation pulse drives the domain wall over the next energy barrier into a neighbouring minimum (inset in Fig. 5c), where it remains trapped after the pulse has vanished. This stepwise accumulation of domain-wall displacements enables the gradual growth of the icicle-shaped protrusion. As the protrusion elongates, the increasing domain-wall tension progressively raises the energy of the distorted configuration until the system approaches an instability. A pulse of opposite polarity then efficiently drives the wall back towards the lower-energy initial state, resulting in the rapid collapse of the protrusion.

The simulations thus demonstrate that the reversible phonon-driven reconfiguration of Néel domains is governed by the interplay between domain-wall elasticity and the local pinning landscape. Rather than simply hindering motion, pinning stabilizes intermediate domain-wall configurations, enabling the accumulation of successive phonon-induced displacements and deterministic switching in selected regions of the crystal.

**Discussion and Outlook**

Our experiments and numerical simulations thus show that the observed antiferromagnetic domain reconfigurations and associated domain wall motions can be interpreted in terms of a helicity-dependent effective field $\mathbf{H}_{\mathrm{L}}^{(\mathrm{eff})}$ acting on the Néel order parameter. For circularly polarized mid-infrared light incident along the sample normal $z$, the effective field is proportional to $E_x E_y^*$, where $E_x$ is the $x$-component of the electric field and $E_y^*$ is the complex conjugate of its $y$-component. For circularly polarized light, this product changes sign upon time reversal[46], similarly to the Néel vector in $MnPS_3$, but unlike the Néel vector it is insensitive to spatial inversion. Hence, a direct thermodynamic coupling between the circularly polarized light and the Néel order in bulk $MnPS_3$ is forbidden.

This restriction can be lifted by the spatial variation of the components of the MIR electric field. The corresponding gradient, $\boldsymbol{\nabla}(E_x E_y^*)$, changes sign under spatial inversion and can therefore provide the additional symmetry required for coupling to the Néel order. For the light incident along the z-axis the most natural spatial variation is associated with its propagation, giving a contribution proportional to $E_x E_y^* k_z$ where $\mathbf{k} = (0,0,k_z)$ is the wavevector of the MIR light. The effective field can be therefore written phenomenologically as:

$$(\mathbf{H}_{\mathrm{L}}^{(\mathrm{eff})})_m = \chi_{mxyz} E_x E_y^* k_z$$

where $\chi_{mijk}$ is a phenomenological fourth-rank tensor present in the *C2/m* space group [47]. In the non-dissipative approximation $\chi_{mijk}$ is purely real, whereas $E_x E_y^*$ is purely imaginary for circularly polarized light. The resulting effective field therefore requires a dissipative contribution to acquire a real component. It is indeed the case in our experiment, where the effect of circularly polarized light is resonantly enhanced at the $A_u$+$B_u$ phonon frequency. Specifically, coherent excitation of two orthogonal, nearly degenerate phonon coordinates, $A_u$ and $B_u$, generates elliptical lattice motion with a well-defined handedness (see Supplementary Information S8 and Table II), thereby producing the net helicity-dependent effective field directly acting on the Néel order.

This phenomenology connects our results to previous studies in which helicity-dependent effective fields were invoked to describe antiferromagnetic control by circularly polarized visible light. In particular, helicity-dependent domain-wall motion[46] and even complete reversal of the Néel order[48,49] have been reported, with the latter described in terms of an "axion"-like effective field in the PT-symmetric van der Waals antiferromagnet $MnBi_2Te_4$[48]. Unlike these studies, where the effective field is associated with electronic excitations, our results show that coherent chiral lattice motion can generate the same symmetry-allowed interaction. At the microscopic level, antiferro-chiral phonons, recently predicted in PT-symmetric antiferromagnets and specifically in $MnPS_3$, may provide a direct coupling between the chiral lattice dynamics and the Néel order[19].

Our results also highlight an important role of the pinning landscape. The observed domain reconfiguration is not driven solely by the phononic field but is strongly facilitated by defect-mediated pinning, which enables the accumulation of small helicity-dependent domain-wall displacements into a stable domain transformation. This suggests that the switching efficiency is governed not only by the magnitude of the effective field but greatly defined the local energy landscape experienced by domain walls. More broadly, our work identifies defect engineering, including the recent approaches based on controlled ion irradiation[50,51], as a promising strategy for tailoring the pinning landscape and optimizing switching in antiferromagnets.

## Acknowledgments

We are grateful to K. Saeedi and C. Berkhout for technical support, and B. A. Ivanov, M. V. Mostovoy and M. X. Na for insightful discussions of the results. This work was supported by the European Research Council through the ERC grants ASTRAL (101078206), HANDSHAKE (101115234), SPARTACUS (101054664); by the Gravitation programme Materials for the Quantum Age (QuMat, 024.005.006), funded by the Dutch Ministry of Education, Culture and Science (OCW); and by the European Union Horizon 2020 Research and Innovation Programme through the Marie Skłodowska-Curie grant COMRAD (861300). We further acknowledge support from the Netherlands Initiative for Energy-Efficient Computing (NL-ECO), part of the Dutch National Science Agenda (NWA-ORC), under grant agreement NWA.1389.20.140.

## Methods

**Sample.** Bulk single crystals of $Mn_{0.9}Ni_{0.1}PS_3$ were grown by chemical vapour transport (CVT), following the procedure of Susner *et al.* [52]. Stoichiometric amounts of Mn and Ni powders, P lumps, and S were sealed in evacuated quartz ampoules ($10^{-5}$–$10^{-6}$ Torr) together with a small amount of halogen transport agent ($I_2$ or $Br_2$, typically 2 mg cm$^{-3}$). Crystal growth was carried out in a two-zone horizontal furnace. To promote complete reaction of the starting materials and suppress parasitic phase formation, the source and growth zones were initially maintained at 650 °C and 750 °C, respectively, for 48 h. Subsequently, the source-zone temperature was increased to 750 °C while the growth-zone temperature was reduced to 700 °C and maintained for 120 h, resulting in the formation of plate-like single crystals. After growth, the furnace was cooled to room temperature and the crystals were recovered for further characterization and experiments. Sample composition and homogeneity were verified by energy-dispersive X-ray spectroscopy (EDS) using a Hitachi SU8020 ultra-high-resolution scanning electron microscope equipped with an Oxford X-MaxN silicon drift detector. Measurements were performed on selected crystals mounted on conductive carbon tape using an acceleration voltage of 20 kV. Additional characterization details are provided in Supplementary Figure 1.

**SHG imaging of 180° Néel domains.** Second-harmonic generation (SHG) microscopy was used to map the spatial distribution of antiferromagnetic domains before and after macropulse exposure. The sample was probed in a wide-field geometry using linearly polarized pulses from a regeneratively amplified Ti laser (800 nm, 25 fs, 1 kHz repetition rate). The generated second-harmonic signal at 400 nm was collected in transmission and detected with a quantitative Complementary Metal-Oxide-Semiconductor camera (q-CMOS, Hamamatsu) after spectral filtering of the residual fundamental beam using a 400±5 nm bandpass filter. (see Supplementary Figure S9)

**MIR FELIX macropulses.** Mid-infrared (MIR) pump pulses were generated by the free-electron laser FELIX (Nijmegen, The Netherlands). The FEL delivers wavelength-tunable picosecond-long micropulses, which are grouped into macropulses consisting of pulse trains. The central wavelength is tunable between 3 and 100 µm, with spectral bandwidths of 0.5–2%. Unless stated otherwise, the macropulse energy at the sample position was approximately 0.2 mJ. The MIR excitation was focused onto the sample using a 90° off-axis parabolic mirror (see Supplementary Figure S9), producing an elliptical focal spot with full-width-at-half-maximum dimensions of 240×230 µm$^2$. In the macropulse experiments, electronic gating of the delay between the arrival of the FEL pulse train and the detection window enabled imaging of the cumulative evolution of the domain reconfiguration induced by successive micropulses within the macropulse.

**Generation of circularly polarized MIR pulses.** The linearly polarized MIR pulses from FELIX were converted into circular polarization using a wavelength-tunable zero-order CdSe quarter-waveplate (Alphalas). The tilt of the quarter-waveplate was adjusted for each pump wavelength to compensate for the wavelength dependence of the retardance. To switch the helicity while avoiding beam displacement caused by rotating the 5-mm-thick CdSe plate at non-normal incidence, the polarization handedness was changed by rotating a KRS-5 wire-grid polarizer placed before the quarter-waveplate. This approach ensured that the propagation direction and spatial profile of the MIR beam remained unchanged upon helicity reversal. The absence of helicity-dependent beam displacement was further confirmed by mapping the pump position through helicity-dependent MIR switching in GdFeCo[45], which enables direct visualization of the excitation spot in the sample focal plane (see Supplementary Figure S10).

**Time-resolved MIR pump-probe stroboscopic SHG experiments.** In the stroboscopic pump–probe experiments (see Supplementary Figure S9), sub-picosecond (≈250 fs) mid-infrared (MIR) pump pulses were generated via difference-frequency mixing in a GaSe crystal using the outputs of two independently tunable optical parametric amplifiers (OPAs) integrated in a single housing (Light Conversion, TOPAS-Twins). The OPAs were pumped by a Ti:sapphire regenerative amplifier (Spectra-Physics Spitfire) operating at a 1 kHz repetition rate, delivering 100 fs pulses at 1.55 eV (800 nm). Both OPAs were seeded by a common white-light continuum generated in a sapphire crystal, ensuring phase-locked outputs with independently tunable wavelengths in the range 1.1–2.7 μm. Mixing of the two OPA outputs in GaSe produced carrier-envelope-phase-stable, linearly polarized MIR pulses tunable between 5 and 19 μm with a typical pulse duration of approximately 250 fs. These pulses were focused onto the sample using a 90° off-axis parabolic mirror to a spot size of approximately 150 μm in diameter. The time-resolved dynamics were probed using a time-delayed optical probe beam controlled by a mechanical delay line in a transmission geometry. The Néel-order dynamics were monitored via second-harmonic generation (SHG), and the SHG intensity was detected using a photomultiplier tube (Hamamatsu H16146-110).

# Supplementary Information (S)

## Deterministic control of antiferromagnetic domain walls by circular phonons

Vladislav Bilyk[1], Nikolai Khokhlov[1], Viktoriia Radovskaia[1], Peter Kim[1], Pietro Diona[2,7], Ravi Kaushik[2,8], Luca Maranzana[2,8], Takeshi Hayashida[3,1], Sergey Artyukhin[2], E. H. T. Teo[4,5], A. Chaturvedi[6], Carl S. Davies[3,1], Andrei Kirilyuk[3,1], Alexey Kimel[1]; Dmytro Afanasiev[1]

**Table of contents:**

**S1. Linear and nonlinear microscopy of the Néel phase**

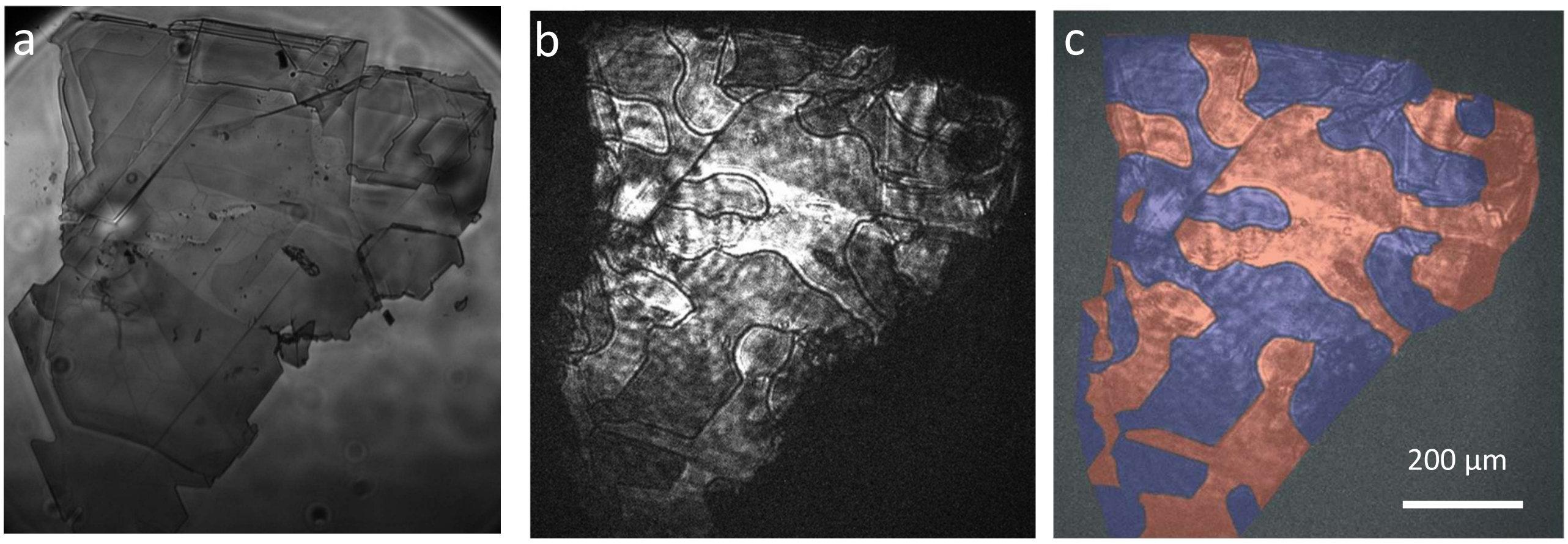

**Figure S1. Optical and SHG microscopy images of $MnPS_3$:Ni on $CaF_2$ substrate**: **(a)** White-light image of the studied sample. **(b)** Second-harmonic generation (SHG) image of the same sample acquired using 800 nm excitation, revealing the antiferromagnetic domain structure. **(c)** False-colour representation of the SHG image, where domains of opposite Néel orientation are shown in red and blue and separated by the domain boundaries.

## S2. Thermal stability of the antiferromagnetic Néel domain pattern

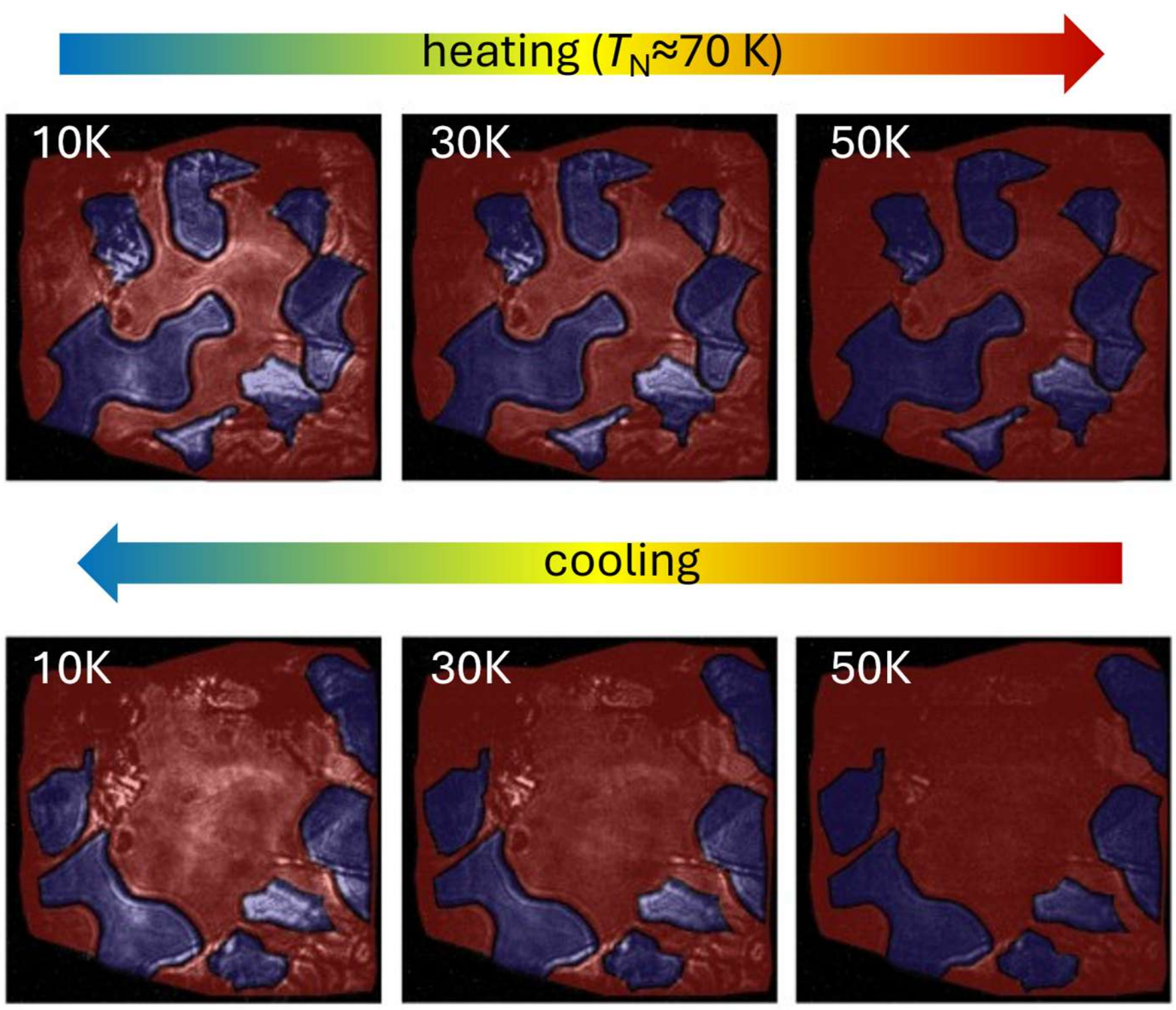

**Figure S2. Thermal cycling of the antiferromagnetic domain pattern.** Microscopic SHG images of the $MnPS_3$:Ni domain structure during heating above $T_N$ and subsequent cooling. The domain pattern remains unchanged up to $T_N$, where the antiferromagnetic order disappears. Upon cooling, a different domain configuration is formed, indicating that thermal cycling through $T_N$ resets the domain structure.

### S3. FTIR absorption and reflection measurements of $MnPS_3$

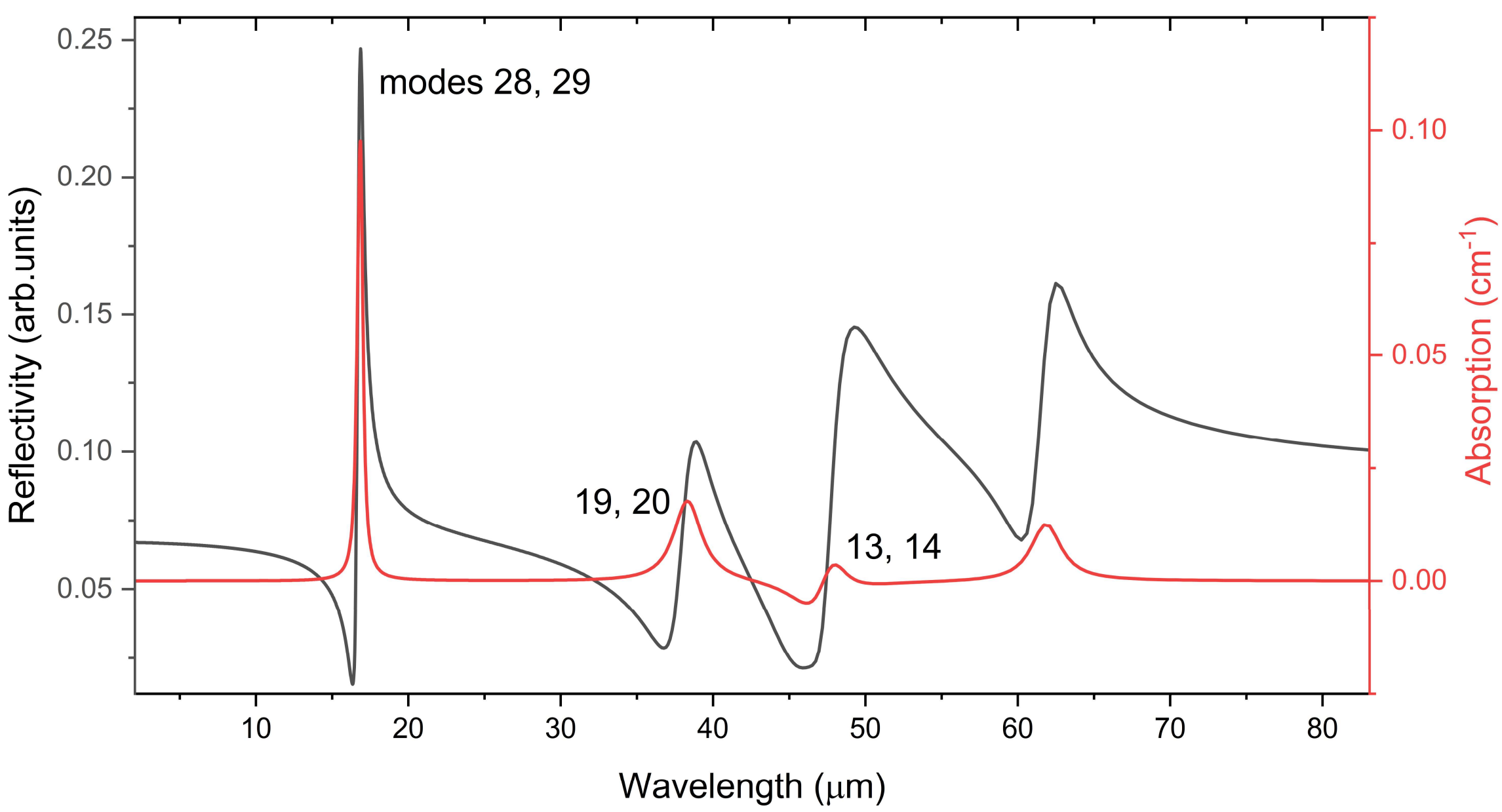

**Figure S3. FTIR measurements** of $MnPS_3$ showing the absorption and reflection spectra over the investigated mid-infrared range. The numbers marked near the peaks correspond to the calculated vibrational mode numbers (see Supplementary Information S8).

**S4. Wavelength dependence of domain reconfiguration on $CaF_2$ and diamond substrates**

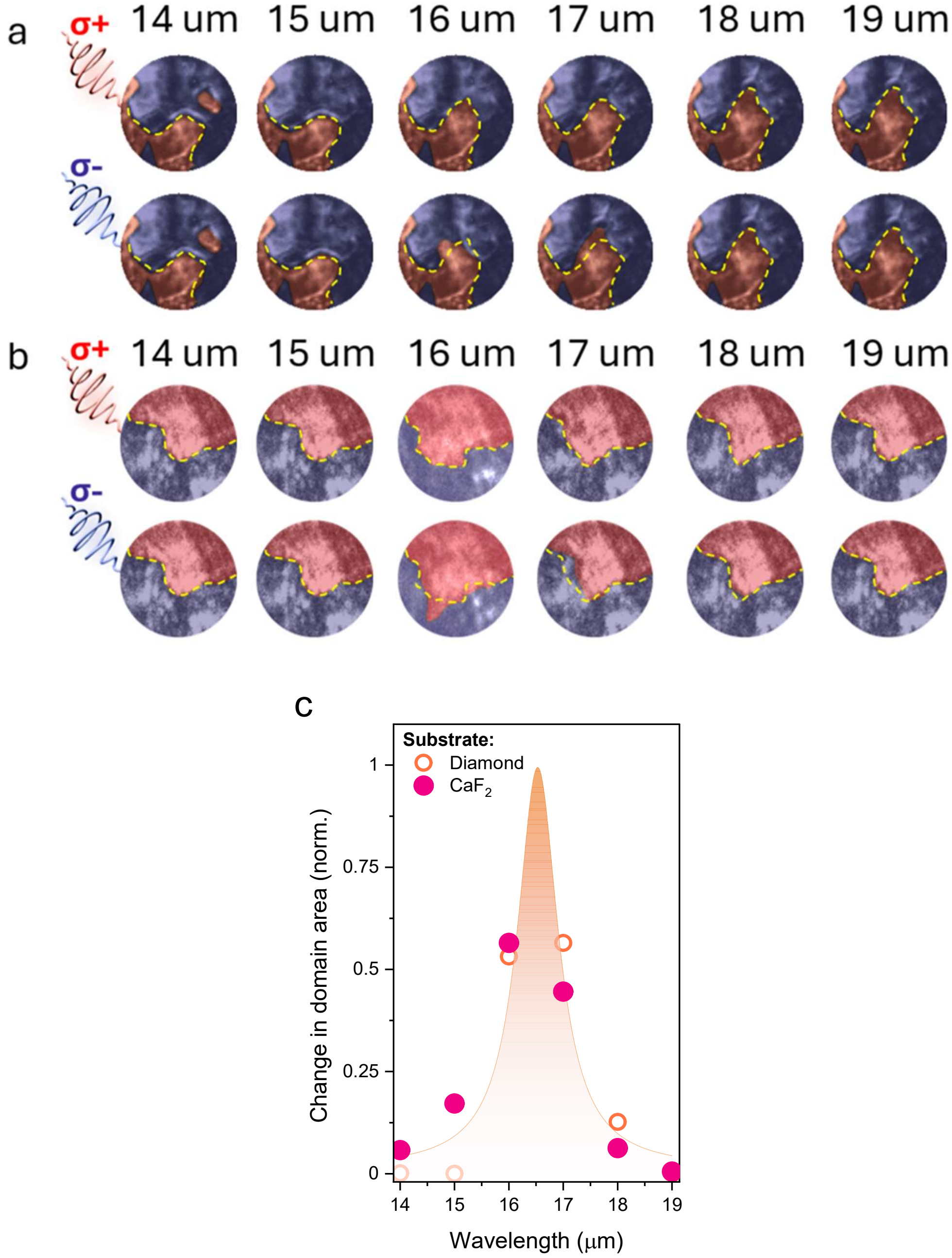

**Figure S4. Spectral dependence of phonon-driven domain reconfiguration.** SHG microscopy images obtained after excitation with right- and left-circularly polarized MIR pulses at different wavelengths for $MnPS_3$:Ni flakes on **(a)** $CaF_2$ and **(b)** diamond substrates. The yellow dashed line marks the domain-wall position after right-circularly polarized excitation. For each wavelength, the sample is first excited with right-circular polarization and then with left-circular polarization for comparison. The wavelength is then tuned and the right- and left-circular pulse sequence is repeated. The domain-wall position after right-circular excitation is marked by the yellow dashed line and retained as a reference in the corresponding left-circular image to highlight the domain reconfiguration. **(c)** Normalized helicity-induced change in the red-domain area as a function of excitation wavelength.

**S5. Helicity-dependent phonon-driven Néel domain reconfiguration at various temepratures**

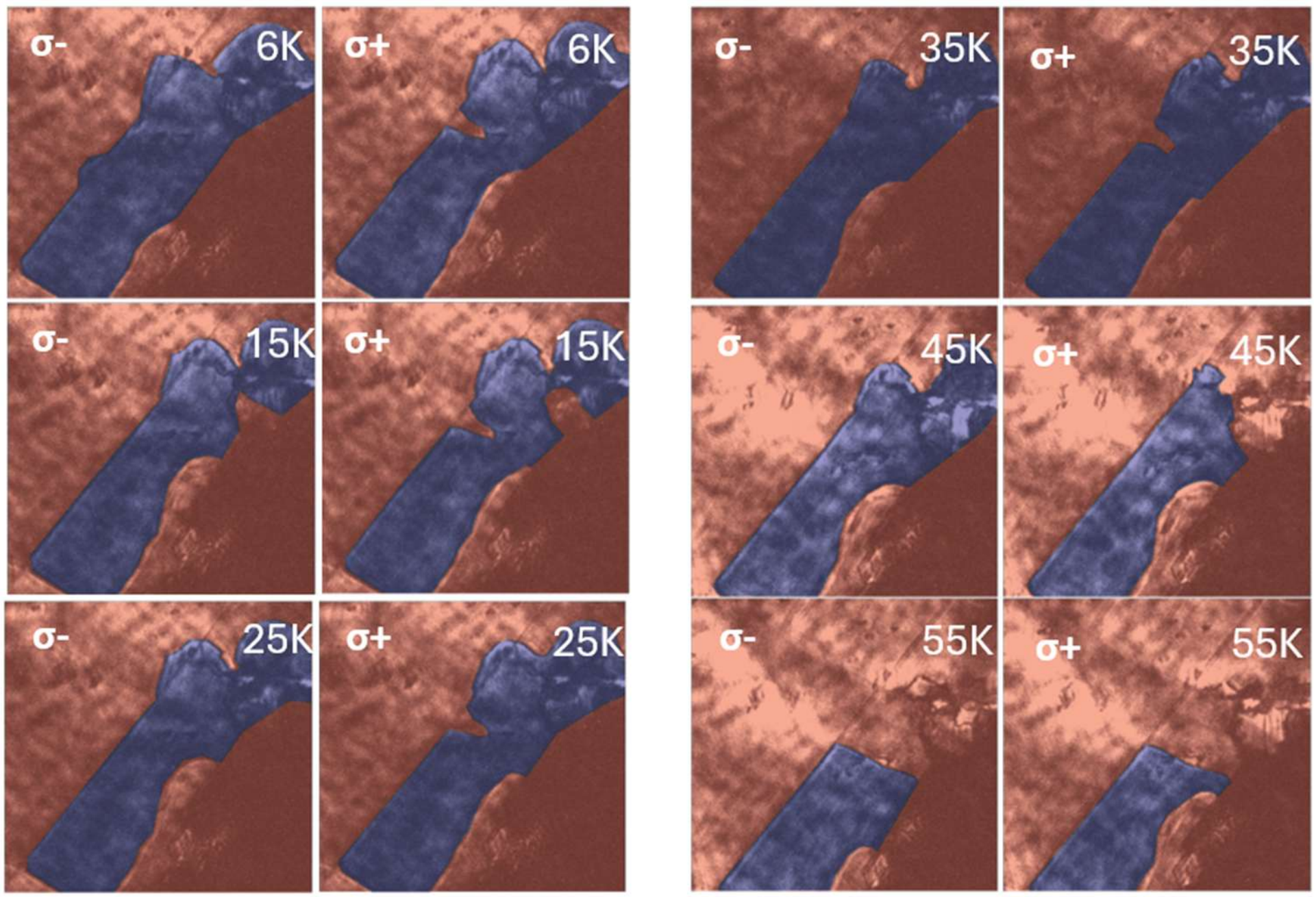

**Figure S5. Full-frame images of the MIR-induced domain reconfiguration shown in Fig. 3b.** Extended SHG microscopy images showing the evolution of the antiferromagnetic domain configuration after a sequence of left- and right-circularly polarized mid-infrared pulses at a selected temperature up to $T_N$. At each temperature, the sample is excited sequentially with left- and right-circular polarization, after which the temperature is changed, and the pulse sequence is repeated.

**S6. Numerical modeling**
Here, we discuss the nature of the asymmetric domain wall motion with respect to laser pulses of opposite circular polarization, and derive a simplified form for the restoring force.

We associate this asymmetry with the elongation $w$ of the wall in the sample's $xy$-plane, that acts as a spring favoring its displacement only in one direction. Therefore, the domain wall energy reads as:

$$E(q) = \mathcal{E} \cdot S(q) = \sqrt{AK} \cdot S(q) = \sqrt{AK}w(q)t,$$

Where $\mathcal{E}$ is domain wall energy per unit area, $A$ is the order parameter $L$ exchange stiffness, $K$ is the easy-axis anisotropy along the $x$-axis, $S$ is the domain wall area, $q$ is the domain wall center and $t$ is the depth of the domain wall along the $z$-axis. From Eq. 3, the geometric force acting on the wall is:

$$F_{\mathrm{DW}} = -\frac{\partial E(q)}{\partial q} = -\sqrt{AK}t\frac{\partial w(q)}{\partial q}.$$

The domain wall shape and position in the $xy$-plane depends on the balancing between three forces: the lattice pinning potential, the effective field on the antiferromagnetic order and the geometric force arising from the elongation of the wall. If the pinning potential together with the geometric elongation is strong enough, the effective field will not be able to displace the domain wall.

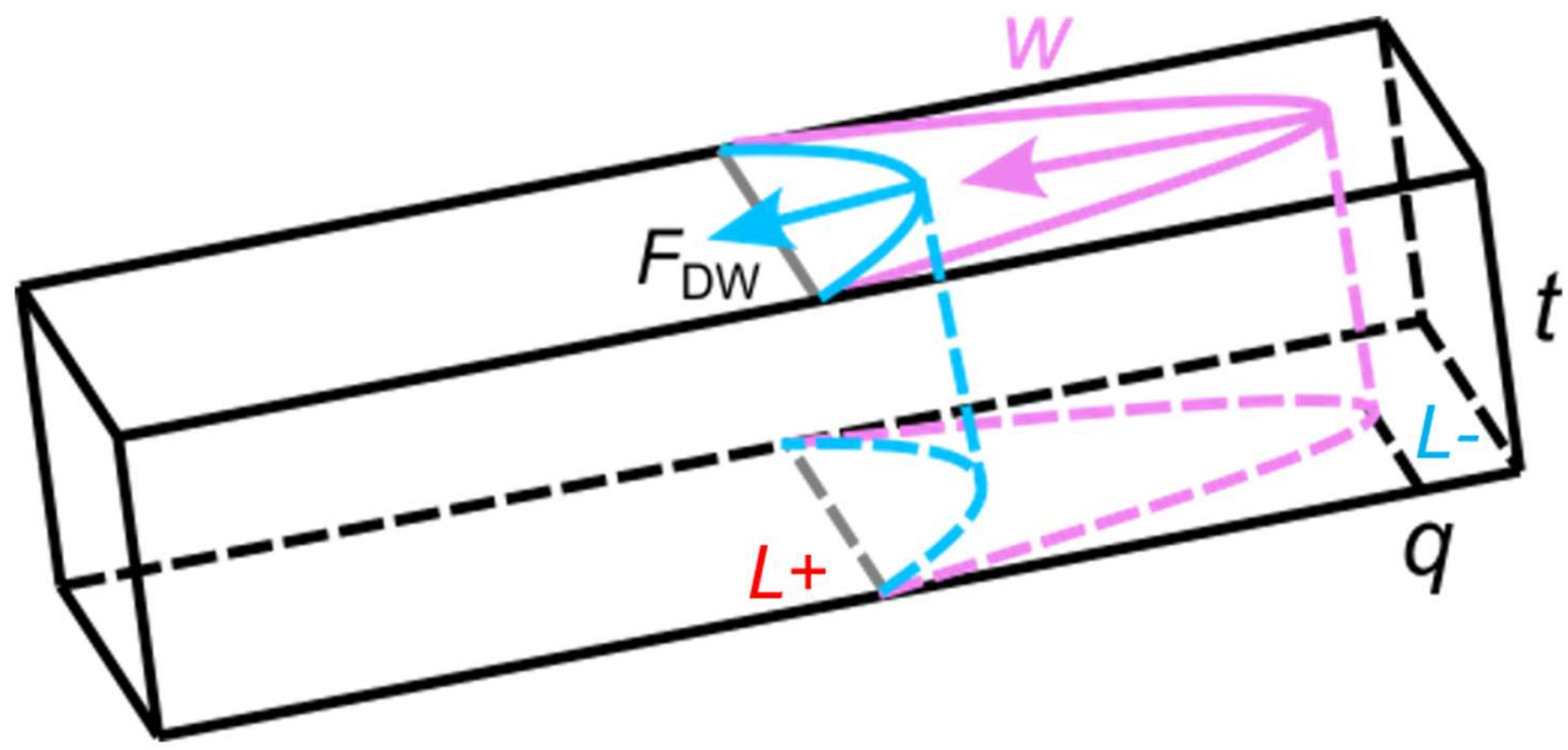

**Figure S6**. **Asymmetric domain wall motion and restoring force.** Schematic illustration of the domain-wall elongation $w$ in the $xy$-plane and the resulting geometric restoring force.

**S7. Micromagnetic simulations**

Micromagnetic simulations are performed with Mumax3. We simulate the order parameter $L$ in a cuboid of 32x32x32 nm$^3$ and lattice spacing 1 nm, with the following Hamiltonian:

$$\mathcal{H} = A\left(\nabla\hat{L}\right)^2 - KL_{\mathrm{x}}^2 - \mathbf{H}_{\mathrm{L}}^{(\mathrm{eff})} \cdot \hat{L}$$

where $\mathbf{H}_{\mathrm{L}}^{(\mathrm{eff})}$ is the effective field generated by the circularity polarized phonon pulse, acting on the antiferromagnetic order. The system is initialized with a 3D domain wall and the pinning sites at the top surface are modeled with a local increase of the exchange interaction.

## S8. Ionic trajectories driven by circularly polarized excitation

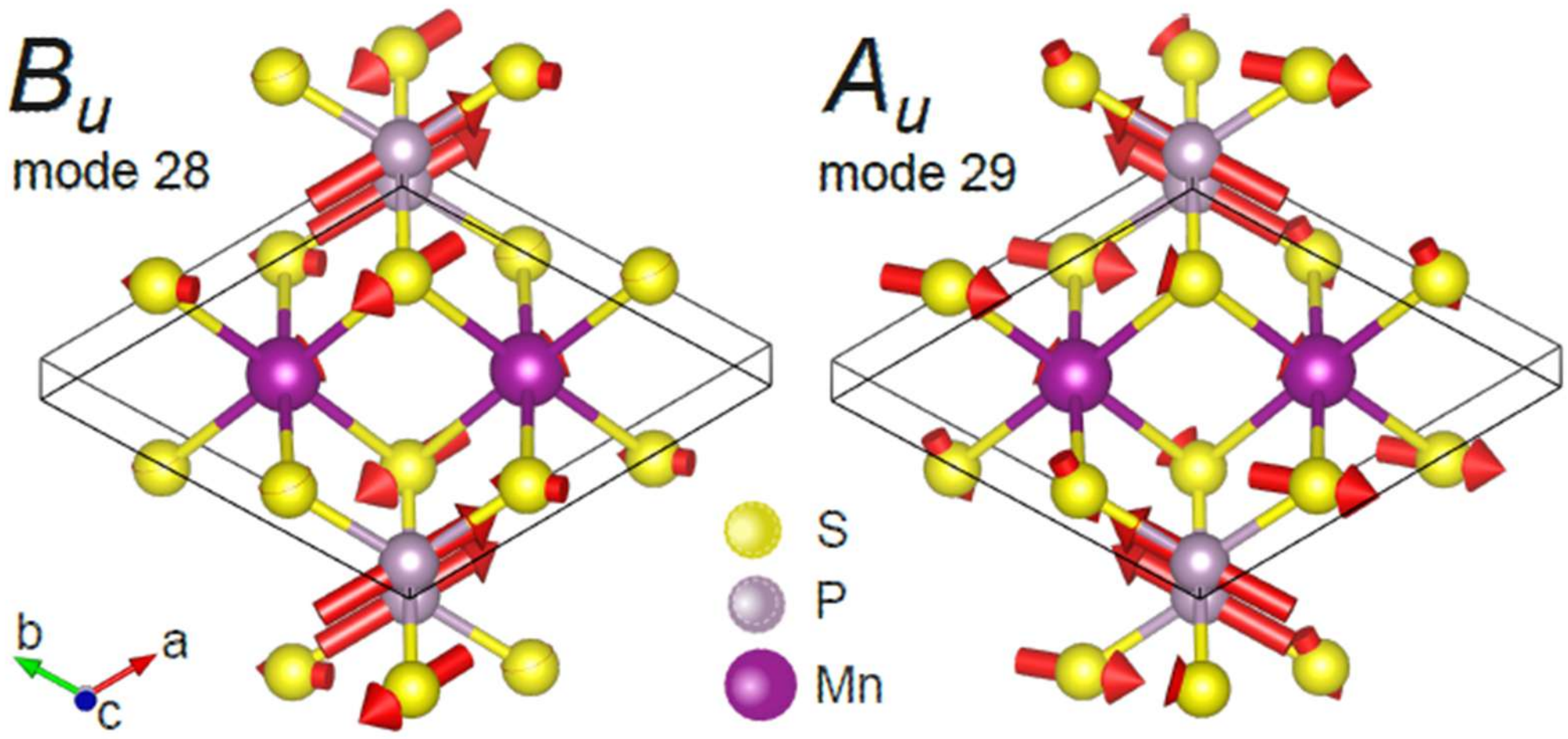

**Figure S8.1**. **Schematic representation of the highest-frequency infrared-active phonon modes.** The circularly polarized MIR light excites the circular motion of atoms, with the perpendicular displacements transforming as $A_u$ and $B_u$ representations. Visualized in VESTA.

The first principles calculations were performed using Quantum ESPRESSO code [1-3]. ONCV pseudopotentials were used [4,5]; the energy cutoff was set to 60 Ry, and the Brillouin zone was discretized on a 6x6x4 *k*-point grid. The phonon calculations were performed using DFPT [6,7] as implemented in Quantum ESPRESSO. The arrows in Fig. S5 show the ionic displacements in the phonon modes, visualized in VESTA [8].

$MnPS_3$ has C2*/m* symmetry group (#12 in the International Tables). Phonons can be classified according to irreducible representations of this group, with their characters indicated in Table SI.

The zone-center vibrational representation is:

$$\Gamma = 8A_g \oplus 7B_g \oplus 6A_u \oplus 9B_u,$$

including the acoustic modes $A_u \oplus 2B_u$. $A_u$ phonons are polarized along the 2-fold axis ($y$) and $B_u$ – perpendicular to it. The optical phonons decompose as:

$$\Gamma_{\text{opt}} = 8A_g \oplus 7B_g \oplus 5A_u \oplus 7B_u.$$

The highest frequency IR active modes are modes 28,29 near 551 $cm^{-1}$; see Table SII and Figure 4. The circularly polarized THz light at this frequency excites the circular motion of atoms, with the perpendicular displacements transforming as $A_u$ and $B_u$ representations. Such circular atomic motion is also supported by $A_u + B_u$ modes near 263 and 201 $cm^{-1}$.

**TABLE SI.** Irreducible representations of the $C_{2h}$ point group classified by the eigenvalues of generators ($C_2, i$).

| $C_2$ $i$ | Irrep |
|---|---|
| +1 +1 | $A_g$ |
| −1 +1 | $B_g$ |
| +1 −1 | $A_u$ |
| −1 −1 | $B_u$ |

**TABLE SII.** Zone-center phonon frequencies, irreducible representations, and optical activity. The modes, excited in our experiment, are highlighted in red.

| Mode number | Frequency ($cm^{-1}$) | Irrep | Activity |
|---|---|---|---|
| 1–3 | 0 | Acoustic ($A_u + 2B_u$) — | |
| 4 | 115.12 | $B_g$ | Raman |
| 5 | 127.96 | $A_g$ | Raman |
| 6 | 130.90 | $B_g$ | Raman |
| 7 | 171.19 | $A_g$ | Raman |
| 8 | 173.26 | $B_g$ | Raman |
| 9 | 177.22 | $A_u$ | IR ($E \parallel y$) |
| 10 | 181.17 | $B_u$ | IR ($E \perp y$) |
| 11 | 189.72 | $B_u$ | IR ($E \perp y$) |
| 12 | 194.25 | $A_u$ | IR ($E \parallel y$) |
| 13 | 201.70 | $A_u$ | IR ($E \parallel y$) |
| 14 | 201.91 | $B_u$ | IR ($E \perp y$) |
| 15 | 235.04 | $B_g$ | Raman |
| 16 | 236.00 | $A_g$ | Raman |
| 17 | 241.79 | $A_g$ | Raman |
| 18 | 245.97 | $B_g$ | Raman |
| 19 | 262.06 | $B_u$ | IR ($E \perp y$) |

| 20 | 264.46 | $A_u$ | IR ($E \parallel y$) |
|---|---|---|---|
| 21 | 270.35 | $B_g$ | Raman |
| 22 | 272.07 | $A_g$ | Raman |
| 23 | 320.33 | $B_u$ | IR ($E \perp y$) |
| 24 | 362.79 | $A_g$ | Raman |
| 25 | 430.13 | $B_u$ | IR ($E \perp y$) |
| 26 | 544.54 | $A_g$ | Raman |
| 27 | 546.10 | $B_g$ | Raman |
| 28 | 550.33 | $B_u$ | IR ($E \perp y$) |
| 29 | 552.43 | $A_u$ | IR ($E \parallel y$) |
| 30 | 564.75 | $A_g$ | Raman |

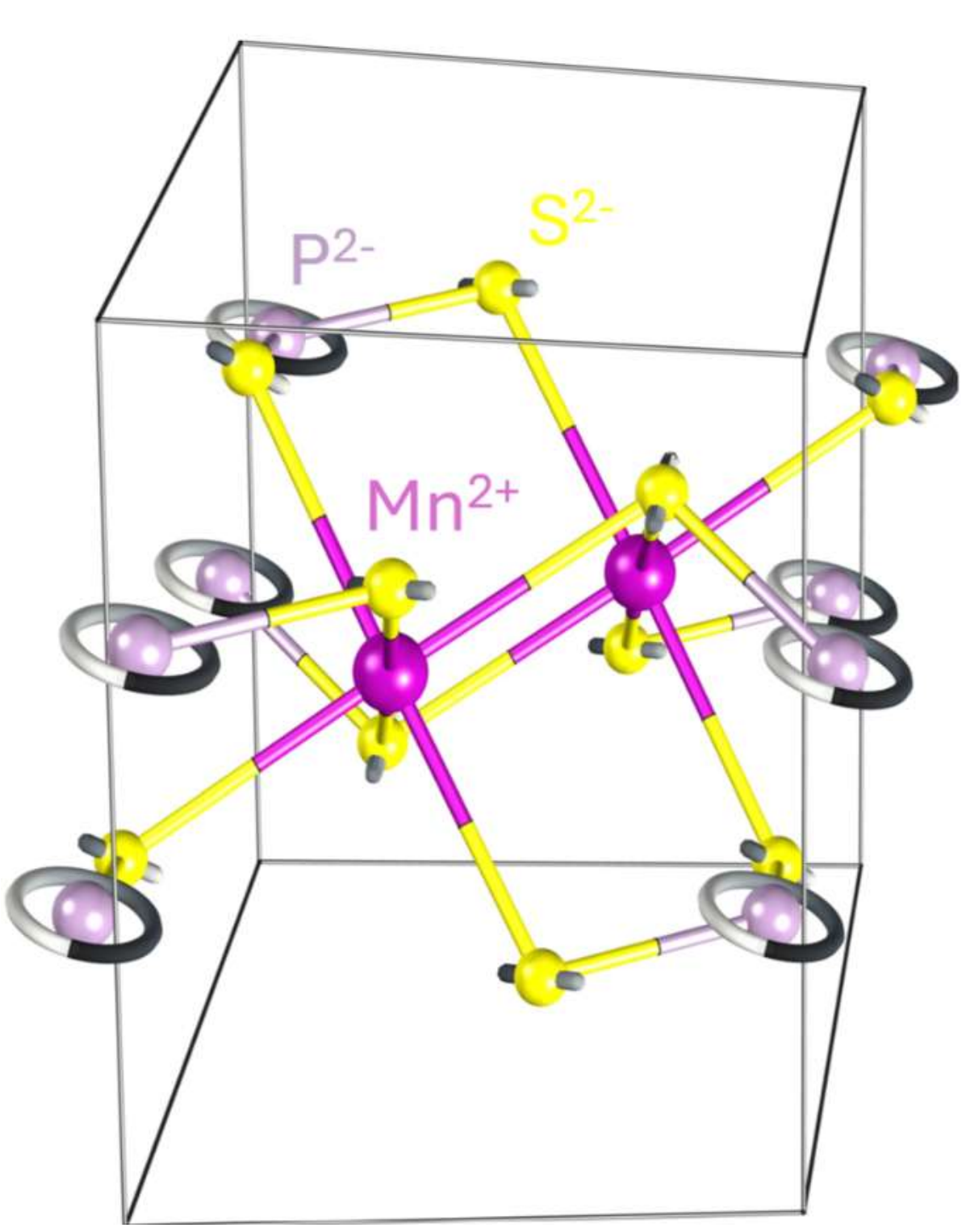

**Figure S8.2. Schematic representation of the ionic trajectories induced by the excitation of the $A_u$, $B_u$ phonons (modes 28, 29 in Table SII).** The circularly polarized MIR light excites the circular motion of atoms, with the perpendicular displacements transforming as $A_u$ and $B_u$ representations.

## S9. Optical layout of the experimental setups

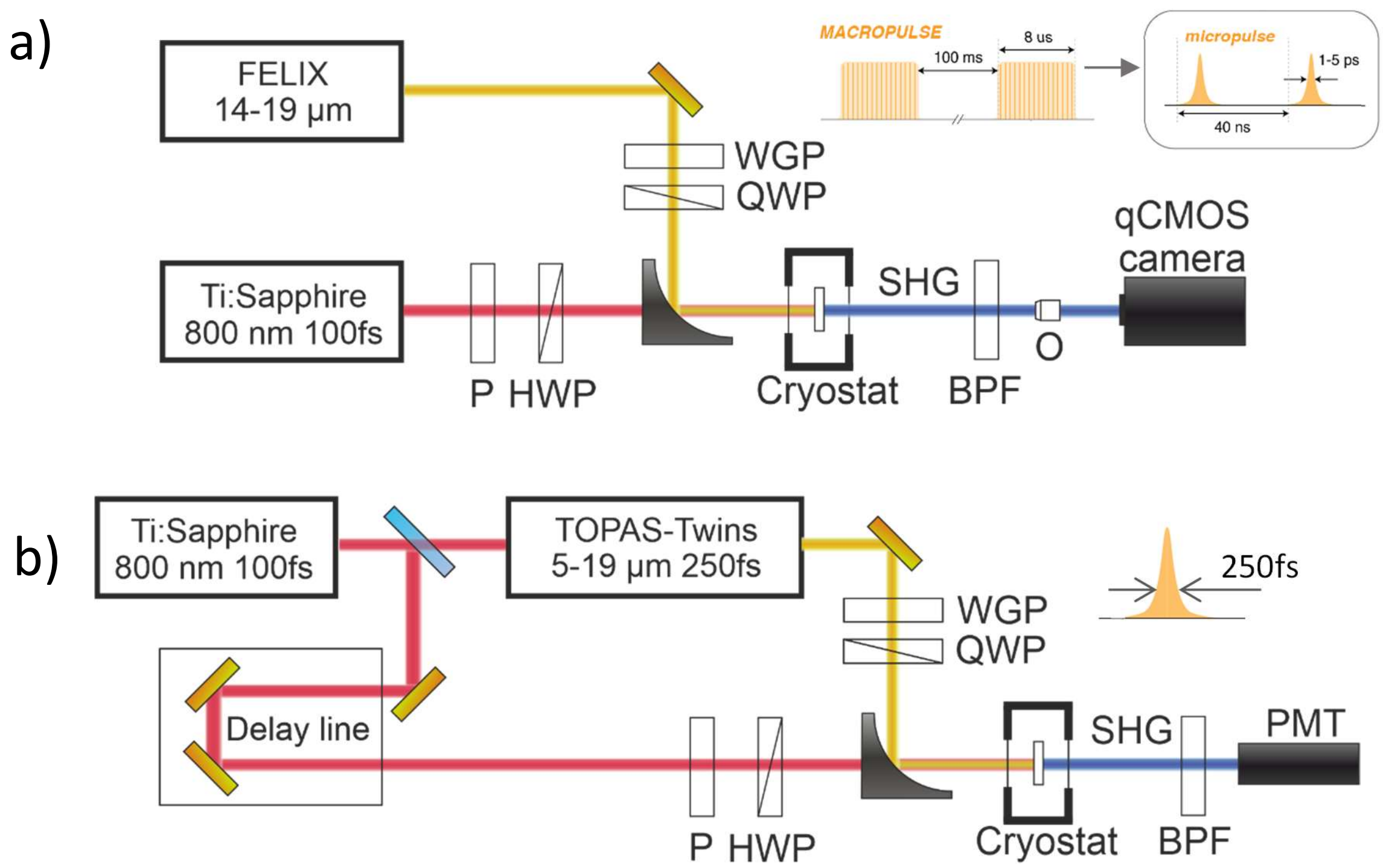

**Figure S9. Schematic representation of the experimental setups.** a) Optical layout for static measurements: FELIX mid-infrared pulses excite the studied sample mounted in a cryostat, while the SHG of the 800 nm beam is used to visualize the antiferromagnetic domain structure. (b) Optical layout for time-resolved measurements: MIR pump-SHG probe configuration used to investigate the ultrafast dynamical response. In both configurations, the MIR beam is focused onto the sample using an off-axis parabolic mirror. A bandpass (BPF) filter is used to suppress the residual 800 nm radiation. WGP: wire-grid polarizer, QWP: quarter-wave plate, P: polarizer, HWP: half-wave plate, PMT: photomultiplier tube.

## S10. Characterization of MIR pump shape and position

To determine the spatial position and profile of the MIR pump excitation, we employ a reference measurement based on ultrafast magnetization reversal in a GdFeCo film deposited on a sapphire substrate. In this system, MIR excitation of infrared-active modes in the substrate induces helicity-dependent switching of the magnetization via the ultrafast Barnett effect. The resulting magneto-optical contrast provides a direct visualization of the pump-induced excitation profile. This enables us to map the spatial position of the MIR pump and confirm the absence of measurable beam displacement upon switching between opposite helicities (see Fig. S4).

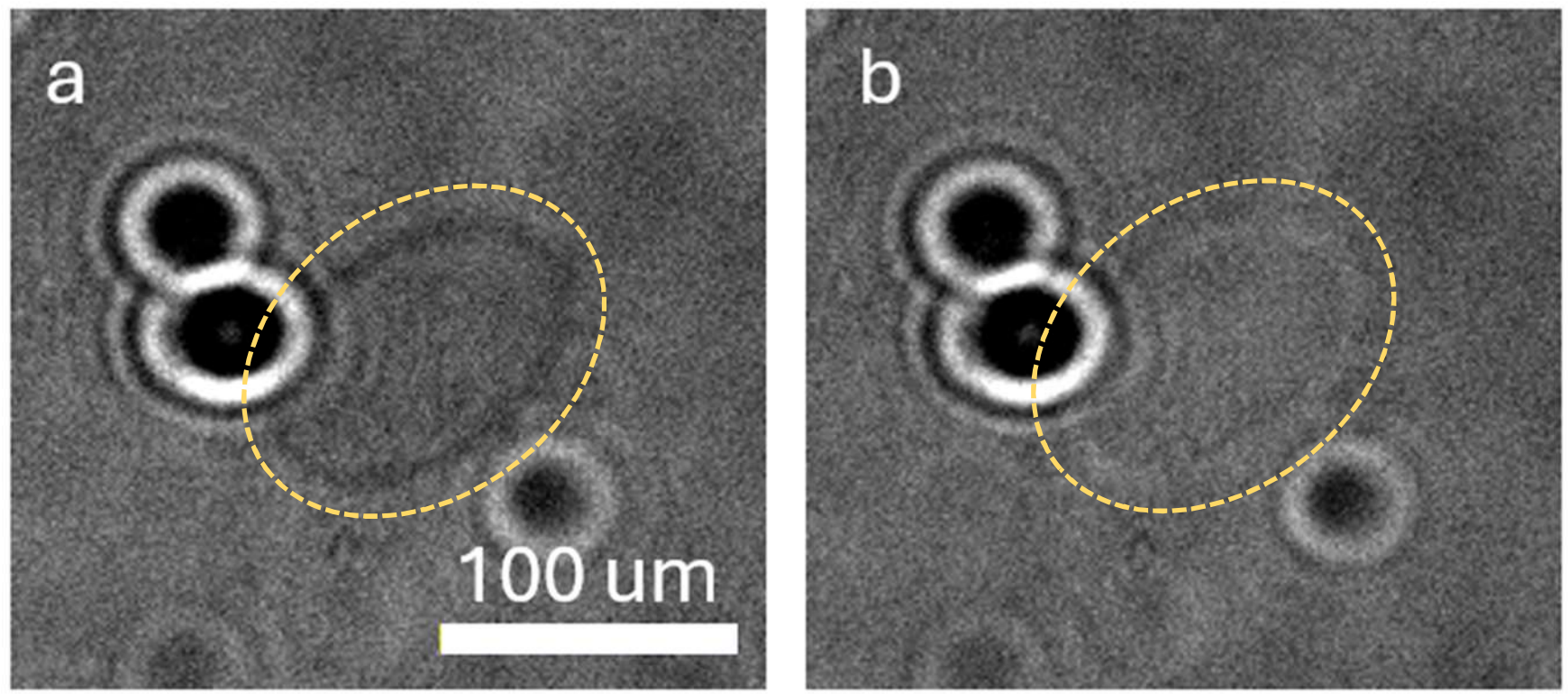

**Figure S10**. **Spatial profile of the MIR pump excitation.** Magneto-optical contrast of a GdFeCo reference film on a sapphire substrate used to visualize the MIR excitation profile via helicity-dependent switching. Images are shown for 16 µm pump excitation with (a) left- and (b) right-circularly polarized pulses. The weak pump-induced contrast appears as dark- and light-grey elliptical regions on the grey background and is highlighted by yellow ellipses as a guide to the eye. The images show no measurable beam displacement upon helicity reversal. Black regions correspond to burned spots on the sample.